\documentclass[prx,twocolumn,floatfix,a4paper,superscriptaddress]{revtex4}

\usepackage{bm,color,amsmath,txfonts}
\usepackage{graphicx}
\usepackage{siunitx}
\usepackage{subfigure}
\usepackage{verbatim}
\usepackage{dcolumn}
\usepackage{bm}
\usepackage{epsf}
\usepackage{xcolor}
\usepackage{hyperref}
\usepackage{hhline}
\usepackage{float}
\usepackage{enumerate}
\usepackage{bbm}
\usepackage{lipsum}
\usepackage{mathrsfs}
\usepackage{ulem}

\begin{document}

\title{Entangling Excitons with Microcavity Photons}

\author{Xuan Zuo}
\affiliation{Interdisciplinary Center of Quantum Information, State Key Laboratory of Modern Optical Instrumentation, and Zhejiang Province Key Laboratory of Quantum Technology and Device, School of Physics, Zhejiang University, Hangzhou 310027, China}
\author{Zhi-Yuan Fan}
\affiliation{Interdisciplinary Center of Quantum Information, State Key Laboratory of Modern Optical Instrumentation, and Zhejiang Province Key Laboratory of Quantum Technology and Device, School of Physics, Zhejiang University, Hangzhou 310027, China}
\author{Hang Qian}
\affiliation{Interdisciplinary Center of Quantum Information, State Key Laboratory of Modern Optical Instrumentation, and Zhejiang Province Key Laboratory of Quantum Technology and Device, School of Physics, Zhejiang University, Hangzhou 310027, China}
\author{Jie Li}\thanks{jieli007@zju.edu.cn}
\affiliation{Interdisciplinary Center of Quantum Information, State Key Laboratory of Modern Optical Instrumentation, and Zhejiang Province Key Laboratory of Quantum Technology and Device, School of Physics, Zhejiang University, Hangzhou 310027, China}

\begin{abstract}
We provide a systemic theory to entangle excitons with microcavity photons. This is realized by adopting an exciton-optomechanics system and introducing a nonlinear dispersive interaction with a mechanical oscillator.  We show that when either the exciton and cavity modes in the weak-coupling regime, or the two exciton-polariton modes in the strong-coupling regime, are respectively resonant with the optomechanical Stokes and anti-Stokes sidebands, entanglement between excitons and cavity photons, or between two exciton polaritons, can be established. The entanglement is in the steady state and can potentially be achievable at room temperature. In both cases, genuine tripartite entanglement is shown to be present.
\end{abstract}

\maketitle

\section{Introduction}

An exciton is an electrically neutral quasiparticle that is formed by the binding of an electron and a hole via the Coulomb interaction. It exists mainly in condensed matter, e.g., insulators and semiconductors, and can interact with electromagnetic fields through the exciton-photon dipole interaction~\cite{Keeling07,Deng10}.  It was theoretically predicted that the strong interaction between excitons and photons can lead to the generation of exciton polaritons~\cite{Hopfield58}, which was first experimentally observed in the semiconductor quantum-well (QW) microcavity~\cite{Weisbuch92,Houdre94}, benefiting from the significantly improved coupling strength between excitons and photons due to the strongly  confined light in the microcavity.   
Since then, extensive research has been made on achieving the exciton-photon strong coupling, e.g., by adopting various microcavity structures, including the micropillar cavity, the photonic crystal slabs, and the whispering gallery microcavity~\cite{Khitrova06,Sun08}, and employing different materials, such as organic~\cite{Lidzey98,Kena-Cohen08}, wide-bandgap~\cite{Sun08}, and perovskite semiconductors~\cite{Su21}. 
Novel exciton-polariton devices have been designed by exploiting the coherence and nonlinearity properties of the polaritons, including spin memory~\cite{Cerna13}, polariton light emitting diodes~\cite{Tsintzos08,Chakraborty19}, and polariton transistors~\cite{Ballarini13,Zasedatelev19}.

As half-matter, half-light bosons, exciton polaritons offer a unique avenue for exploring the interface between quantum optics, spontaneous coherence, and quantum condensation~\cite{Keeling07}. They can be manipulated and probed via the light component and generate rich nonlinear interactions through the matter component~\cite{Keeling07}. Because the effective mass is much lower than that of typical atomic systems, the critical temperature for achieving the Bose-Einstein condensation (BEC) of the exciton polaritons can be several orders of magnitude higher than the atomic one, which is potentially attainable even at room temperature~\cite{Deng10}.  Despite the highly dissipative two-dimensional system with weak particle-particle interactions, which do not conform to the ideal BEC, the exciton-polariton BEC has been successfully achieved in the experiments~\cite{BEC1,BEC2}, with the observation of relevant macroscopic quantum phenomena, including Bogoliubov excitations~\cite{Bogoliubov}, quantized vortices~\cite{vortices}, and superfluidity~\cite{fluid}.  The coherent nature of the polariton condensates can be exploited to create a high-efficiency low-threshold laser without inversion~\cite{Imamoglu96}, which was first demonstrated in the experiments~\cite{Malpuech02,Deng03}. After that, a variety of configurations have been adopted to realize the room-temperature lasing~\cite{Christopoulos07,Bajoni08,Kena-Cohen10}, and a more efficient polariton laser has been demonstrated by using an electrical pump method~\cite{Schneider13,Bhattacharya13,Bhattacharya14}.   In addition, quantum entanglement of the exciton polaritons has been studied by exploiting the parametric process~\cite{Ciuti04,Liew18,DS} and spin squeezing~\cite{Feng}, and the photon-polariton entanglement has been experimentally observed 
by swapping a photon for a polariton in two-photon entangled states~\cite{Cuevas}.  Besides, squeezed light has been produced utilizing the strong nonlinearity of the polariton-polariton interaction~\cite{Boulier14}.

Here, we provide a complete theory to entangle excitons with microcavity photons by coupling the latter to a mechanical oscillator via a nonlinear dispersive interaction. The system then becomes a tripartite bosonic system, namely, exciton-optomechanics (EOM)~\cite{Santos,EOM1,EOM2,EOM3,Fainstein20,Santos21,Santos23}. Due to the linear excitation-exchange (beam-splitter-type) interaction between excitons and photons, the two modes are intrinsically not entangled. The mechanical oscillator couples to the optical cavity mode via a dispersive manner, which can enable an effective optomechanical parametric down-conversion (PDC) or state-swap interaction, corresponding to the optomechanical Stokes or anti-Stokes scattering.   We consider a complete set of two situations depending on whether excitons and microcavity photons are strongly coupled to form exciton polaritons. We show that when either the exciton and cavity modes (in the weak-coupling regime) or the two exciton-polariton modes (in the strong-coupling regime) are respectively resonant with the optomechanical Stokes and anti-Stokes sidebands, the entanglement between excitons and cavity photons, or between two polariton modes, can be established. The entanglement is in the steady state and can be achieved even at room temperature for a not very high mechanical quality factor. We further show that both the excitons and cavity photons (or the two polaritons) are entangled with the mechanical oscillator, exhibiting genuine tripartite entanglement in the EOM system.

The paper is organized as follows. In Sec.~\ref{model1}, we describe the EOM system, provide its Hamiltonian and the corresponding Langevin equations, and show how the steady-state solutions of the system can be achieved. We then present in Sec.~\ref{En1} the results of the exciton-photon entanglement and the tripartite entanglement in the weak-coupling case. In Sec.~\ref{model2}, we reformulate the theory for the exciton-polariton-mechanics system in the strong-coupling regime, and show the results of the entanglement of two polaritons and the tripartite entanglement in Sec.~\ref{En2}.  Finally, we make a discussion on the two entanglement protocols in the weak- and strong-coupling regimes and summarize the findings in Sec.~\ref{conc}.

\begin{figure}[b]
	\includegraphics[width=0.93\linewidth]{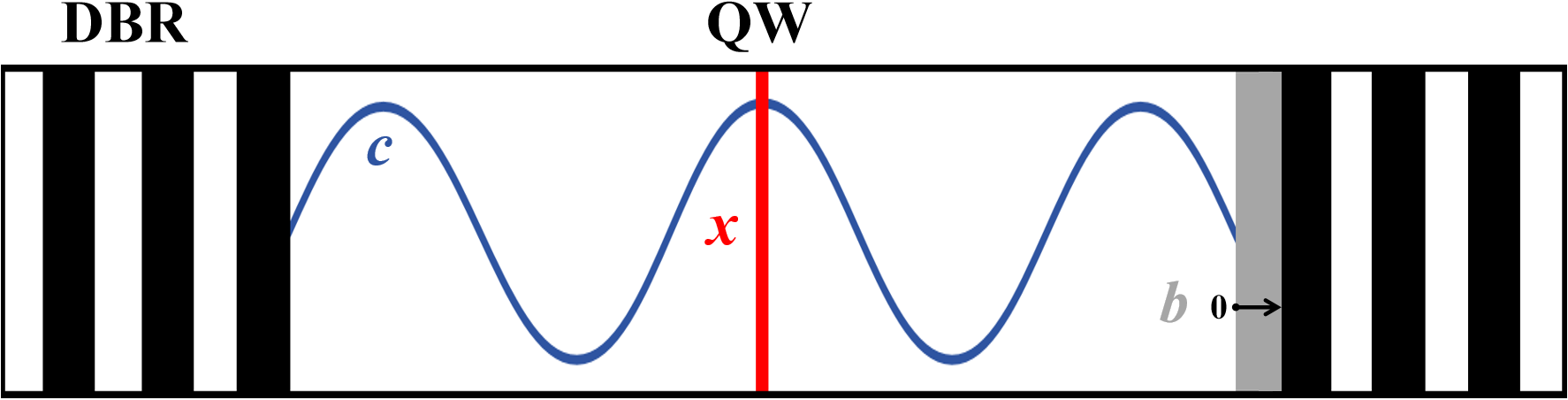}
	\caption{Sketch of the EOM system. A QW is placed within a semiconductor microcavity that is formed by two DBRs, which enables the interaction between the exciton mode ($x$) and the optical cavity mode ($c$).  The DBRs are movable and the mechanical motion ($b$) couples to the cavity photons via a dispersive interaction. The gray area is a diagram of the mechanical motion, which does not correspond to the real motion. 
	}
	\label{fig1}
\end{figure}

\section{The exciton-optomechanics system}\label{model1}

The EOM system consists of a QW and a semiconductor microcavity formed by two movable distributed Bragg reflectors (DBRs), as depicted in Fig.~\ref{fig1}. A DBR is made of layers of alternating high and low refraction indices, and each layer has an optical thickness of $\lambda/4$, with $\lambda$ being the optical wavelength. Thus, light reflected from each interface destructively interfere and create a stop band for transmission. The DBR then acts as a high-reflectivity mirror when the wavelength of the incident light is within the stop band.  The DBRs are movable and the mechanical displacement couples to the cavity photons via a dispersive optomechanical interaction~\cite{EOM1,EOM3,Perrin13}. A QW is a thin layer of semiconductor, which is sandwiched between two barrier layers with a much larger band gap. The QW is placed within the microcavity, which enables a linear excitation-exchange (beam-splitter type) interaction between excitons and cavity photons. The Hamiltonian of the EOM system reads
\begin{align}\label{HHH}
\begin{split}
H/\hbar &= \! \omega_{x} x^\dagger x + \omega_{c} c^\dagger c + \omega_b b^\dagger b + G_0 c^\dagger c \left( b + b^\dagger \right)
\\& + g \left(x^\dagger c + x c^\dagger \right) + i\Omega \left(c^\dagger e^{-i\omega_0t}-c e^{i\omega_0t} \right) ,
\end{split}
\end{align}
where $x$, $c$, and $b$ ($x^\dagger$, $c^\dagger$, and $b^\dagger$) are the annihilation (creation) operators of the excitons, cavity photons, and phonons, respectively, satisfying the commutation relation $[j,j^{\dag}]=1$ ($j=x,\,c,\,b$), and $\omega_x$, $\omega_c$, and $\omega_b$ are their resonance frequencies. Here, $g$ denotes the exciton-photon coupling strength, which is variable depending on the position of the QW placed within the microcavity and can be very strong. When the coupling strength exceeds the exciton and cavity decay rates $\kappa_x$ and $\kappa_c$, i.e., $g>\kappa_x,\kappa_c$, the system enters the strong-coupling regime leading to the exciton polaritons~\cite{Weisbuch92,Houdre94}.  The single-photon optomechanical coupling strength $G_0$ is typically weak~\cite{MA14}, but the effective optomechanical coupling can be significantly enhanced by driving the microcavity with an intense laser field. The last term corresponds to the driving Hamiltonian, where $\Omega = \sqrt{2 P \kappa_c/ \hbar \omega_0}$ signifies the coupling strength between the cavity and the drive field with frequency $\omega_0$ and power $P$.

By incorporating the dissipation and input noise of each mode, we obtain the following quantum Langevin equations (QLEs) in the frame rotating at the drive frequency $\omega_0$:
\begin{align}\label{QLExcb}
	\begin{split}
		\dot{x}=&-(i\Delta_x + \kappa_x) x - i g c + \sqrt[]{2\kappa_x}x^{in},  \\
		\dot{c}=&-(i\Delta_c + \kappa_c) c - i g x - i G_0 c ( b + b^\dagger ) + \Omega + \sqrt[]{2\kappa_c}c^{in}, \\
		\dot{b}=&-(i\omega_b + \kappa_b) b - i G_0 c^\dagger c + \sqrt[]{2\kappa_b} b^{in} ,
	\end{split}
\end{align}
where $\Delta_x=\omega_x-\omega_0$, $\Delta_c=\omega_c-\omega_0$, $\kappa_b$ is the mechanical damping rate, and $j^{in}(t)$ ($j=x,\,c,\,b$) are the input noise operators of the three modes, which are zero-mean and characterized by the correlation functions~\cite{Zoller}: $\langle j^{in}(t)j^{in\dagger}(t^\prime) \rangle=[N_j(\omega_j)+1]\delta(t-t^\prime)$, $\langle j^{in\dagger}(t)j^{in}(t^\prime) \rangle=N_j(\omega_j)\delta(t-t^\prime)$, with $N_j(\omega_j)=[\exp[(\hbar \omega_j/k_BT)]-1]^{-1}$ being the equilibrium mean thermal excitation number of the mode $j$, and $T$ as the bath temperature.

Owing to the strong driving of the cavity and the exciton-photon excitation-exchange interaction, the cavity and exciton modes exhibit large amplitudes $| \langle c \rangle|,\, | \langle x \rangle| \gg 1$. This allows us to linearize the system dynamics around the steady-state values by expressing each mode operator $j$ as the sum of its classical average $\langle j \rangle$ and quantum fluctuation operator $\delta j$, i.e., $j = \langle j \rangle + \delta j$, and neglecting small second-order fluctuation terms. As a result, the QLEs are separated into two sets of equations, one for the classical averages and the other for the quantum fluctuations. The linearized QLEs of the quantum fluctuations are obtained as
\begin{align}\label{QLEs}
	\begin{split}
		\dot{\delta x} = & - \big( i {\Delta}_x + \kappa_x \big) \delta x - i g {\delta c} + \sqrt[]{2\kappa_x} x^{in},  \\
		\dot{\delta c} = & - \big( i \tilde{\Delta}_c + \kappa_c \big) \delta c - i g {\delta x} - G_{cb} \big( \delta b + \delta b^\dagger \big) + \sqrt[]{2\kappa_c} c^{in},  \\
		\dot{\delta b} = & - \big( i {\omega}_b + \kappa_b \big) \delta b -  \left(G_{cb} \delta c^\dagger - G_{cb}^* \delta c \right) + \sqrt[]{2\kappa_b} b^{in},
	\end{split}
\end{align}
where $\tilde{\Delta}_c = {\Delta}_c + 2 G_0 \langle b \rangle$ is the effective cavity-drive detuning including the frequency shift due to the optomechanical interaction, and $G_{cb} = i G_0 \langle c \rangle$ is the effective optomechanical coupling strength. The expressions of the classical averages are given by
\begin{align}\label{stSol}
	\begin{split}
		\langle x \rangle =& \frac{- i \Omega g}{g^2 + (i \tilde{\Delta}_c + \kappa_c) (i {\Delta}_x + \kappa_x)}, \\	 
		\langle c \rangle  =& \frac{\Omega (i {\Delta}_x + \kappa_x)}{g^2 + (i \tilde{\Delta}_c + \kappa_c) (i {\Delta}_x + \kappa_x)}, \\
		\langle b \rangle =& -\frac{G_0}{\omega_b} | \langle c \rangle |^2.
	\end{split}
\end{align}
The QLEs~\eqref{QLEs} can be expressed in a compact matrix form with the quadrature fluctuation operators $\delta X_j = (\delta j + \delta j^\dagger)/\!\sqrt{2}$, and $\delta Y_j = i (\delta j^\dagger - \delta j)/\!\sqrt{2}$, i.e.,
\begin{align}\label{uAn}
\dot{u}(t)= {\cal A} \, u(t) + n(t),
\end{align}
where $u(t)=\left[\delta X_x(t),\delta Y_x(t),\delta X_c(t),\delta Y_c(t),\delta X_b(t),\delta Y_b(t) \right]^{\rm T}$ represents the vector of quantum fluctuations, $n(t) = \left[\sqrt[]{2\kappa_x}X_x^{in},\sqrt[]{2\kappa_x}Y_x^{in},\sqrt[]{2\kappa_c}X_c^{in},\sqrt[]{2\kappa_c}Y_c^{in}, \sqrt[]{2\kappa_b}X_b^{in}, \sqrt[]{2\kappa_b}Y_b^{in} \right]^{\rm T}$ is the vector of input noises, where $X_j^{in}$ and $Y_j^{in}$ are defined similarly as $\delta X_j$ and $\delta Y_j$ but with the noise operators $j^{in}$ and $j^{in\dag}$, and the drift matrix ${\cal A}$ is given by
\begin{align}
	\cal A=\begin{pmatrix}
		-\kappa_x & {\Delta}_x & 0 & g & 0 & 0 \\
		-{\Delta}_x & -\kappa_x & -g & 0 & 0 & 0\\
		0 & g & -\kappa_c & \tilde{\Delta}_c & - 2 {\rm Re}\,G_{cb} &0 \\
		-g & 0 & -\tilde{\Delta}_c & -\kappa_c & - 2 {\rm Im}\,G_{cb} & 0\\
		0 & 0 & 0 & 0 & -\kappa_b & \omega_b \\
		0 & 0 & - 2 {\rm Im}\,G_{cb} & 2 {\rm Re}\,G_{cb} & -\omega_b & -\kappa_b\\
	\end{pmatrix}.
\end{align}
Due to the linearized dynamics and the inherent Gaussian properties of the quantum noises, the steady state of the quadrature fluctuations is a three-mode Gaussian state, which is fully characterized by a $6 \times 6$ covariance matrix (CM) $V$, with its entries $V_{ij}=\frac{1}{2} \langle u_i(t)u_j(t^\prime) + u_j(t^\prime)u_i(t) \rangle$ $(i,j=1,2,...,6)$. The steady-state CM can be obtained by directly solving the Lyapunov equation~\cite{Parks,Vitali07}
\begin{align}
	\begin{split}
	{\cal A} V + V {\cal A}^{\rm T} = -D,
	\end{split}
\end{align}
where $D=\mathrm{Diag} \big[\kappa_x(2N_x + 1),\kappa_x(2N_x + 1),\kappa_c(2N_c + 1),\kappa_c(2N_c + 1),\kappa_b(2N_b+1),\kappa_b(2N_b+1)\big]$ is the diffusion matrix, with its entries defined via $D_{ij}\,\delta(t - t') = \langle n_i(t)n_j(t')+n_j(t')n_i(t) \rangle/2$.

{
Once the CM $V$ is achieved, we adopt the logarithmic negativity $E_N$~\cite{Ent} to quantify the quantum entanglement between any two modes of the system, 
which is defined as
\begin{align}\label{En}
	\begin{split}
	E_N \equiv \max[0,-\ln 2 \tilde{\nu}_-],
	\end{split}
\end{align}
where $\tilde{\nu}_- = \min {\rm eig}|i \Omega_2 \tilde{V}_{\rm 4}|$ (the symplectic matrix $\Omega_2 = \oplus_{j=1}^2 i \sigma_y$ and $\sigma_y$ is the $y$-Pauli matrix) is the minimum symplectic eigenvalue of the CM $\tilde{V}_{\rm 4} = {\cal P} {V}_{\rm 4} {\cal P}$, with ${V}_{\rm 4}$ being the $4 \times 4$ CM of the two modes under consideration, obtained by removing in $V$ the rows and columns associated with the uninteresting modes, and ${\cal P} = \mathrm{Diag} [1, -1, 1, 1]$ being the matrix that performs partial transposition on the CM~\cite{Simon}.}

{
To determine the tripartite entanglement, we adopt the residual contangle~\cite{Cont,Adesso}
\begin{align}
	\begin{split}
	R^{i|jk}_\tau \equiv C_{i|jk} - C_{i|j} - C_{i|k},
	\end{split}
\end{align}
where $C_{u|v}$ is the contangle of subsystems of $u$ and $v$ ($v$ contains one or two modes), which is a proper entanglement monotone defined as the squared logarithmic negativity. To calculate the one-vs-two-modes logarithmic negativity, one only needs to follow the definition of Eq.~\eqref{En} by simply replacing $\Omega_2 = \oplus_{j=1}^2 i \sigma_y$ with $\Omega_3 = \oplus_{j=1}^3 i \sigma_y$, and $\tilde{V}_{\rm 4} = {\cal P} {V}_{\rm 4} {\cal P}$ with $\tilde{V} = {\cal P}_{i|jk} {V} {\cal P}_{i|jk}$, where the partial transposition matrices ${\cal P}_{1|23} = \mathrm{Diag} [1, -1, 1, 1, 1, 1]$, ${\cal P}_{2|13} = \mathrm{Diag} [1, 1, 1, -1, 1, 1]$, and ${\cal P}_{3|12} = \mathrm{Diag} [1, 1, 1, 1, 1, -1]$. The residual contangle satisfies the monogamy of quantum entanglement, $R^{i|jk}_\tau \geq 0$, i.e., 
\begin{align}
	\begin{split}
	C_{i|jk} \geq C_{i|j} + C_{i|k},
	\end{split}
\end{align}
which is similar to the Coffman-Kundu-Wootters monogamy inequality~\cite{Coffman} hold for the system of three qubits. A {\it bona fide} quantification of the tripartite entanglement is given by the minimum residual contangle~\cite{Cont,Adesso}
\begin{align}
	\begin{split}
	R^{\rm min}_\tau \equiv \min[R^{i|jk}_\tau, R^{j|ik}_\tau, R^{k|ij}_\tau],
	\end{split}
\end{align}
which guarantees that $R^{\rm min}_\tau$ is invariant under all permutations of the modes, thus indicating a genuine three-way property of any three-mode Gaussian state. A nonzero minimum residual contangle $R^{\rm min}_\tau>0$ denotes the presence of tripartite entanglement in the system.}

\begin{figure}[t]
	\includegraphics[width=0.93\linewidth]{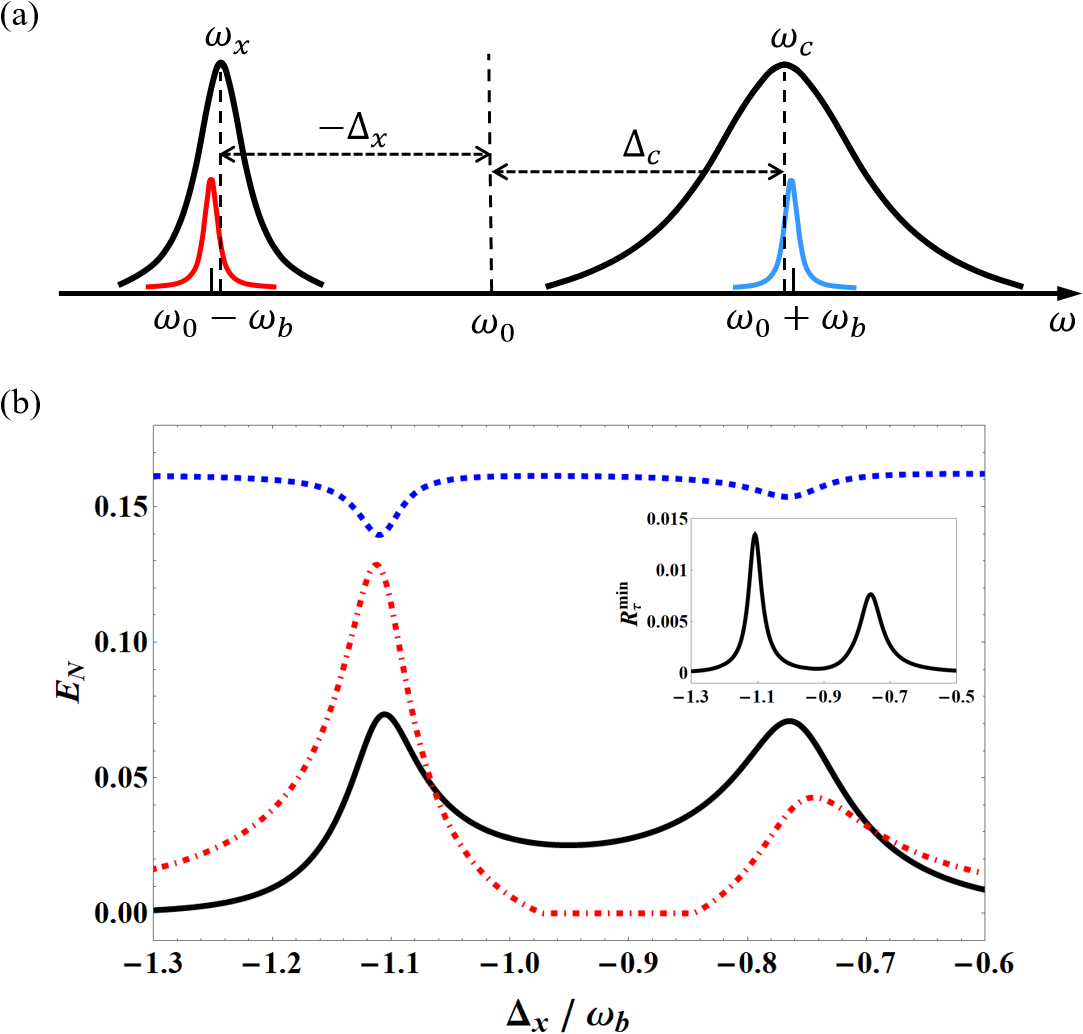}
	\caption{(a) The mechanical motion with frequency $\omega_b$ scatters the driving photons at frequency $\omega_0$ onto two sidebands at $\omega_0 \pm \omega_b$. When the exciton and cavity modes are resonant with the Stokes (red) and anti-Stokes (blue) sidebands, respectively, genuine tripartite entanglement of cavity photons, excitons, and phonons is present. (b) Stationary exciton-photon (solid), photon-phonon (dashed), and exciton-phonon (dot-dashed) entanglement and tripartite entanglement (inset) versus the exciton-drive detuning $\Delta_x$. We take $\tilde{\Delta}_c = 0.9 \omega_b$, and the other parameters are provided in the text.}
	\label{fig2}
\end{figure}

\section{Entanglement between Excitons and Microcavity Photons}\label{En1}

In this section, we consider the weak-coupling case where the interaction between excitons and microcavity photons does not lead to exciton polaritons. Specifically, we consider a relatively weak exciton-photon coupling $g < \kappa_c,\kappa_x$, and the exciton and cavity modes are non-resonant but of close frequencies, cf. Fig.~\ref{fig2}(a).  

The prerequisite for obtaining entanglement in the system is to cool the lower-frequency mechanical mode close to its quantum ground state and eliminate the detrimental thermal noise. To this end, we adopt a red-detuned laser field to drive the microcavity (with $\tilde{\Delta}_c \approx \omega_b \gg \kappa_c$) to activate the optomechanical anti-Stokes scattering, which is responsible for cooling the mechanical motion. For mechanical cooling, the drive field should be relatively weak, such that the rotating-wave approximation (under the weak-coupling condition $G_{cb} \ll \omega_b$) can be taken to obtain the effective optomechanical beam-splitter interaction $\propto \delta{c}^\dagger \delta{b} + \delta{c} \delta{b}^\dagger$~\cite{DV08}.  However, the optomechanical beam-splitter interaction, as well as the linear exciton-photon interaction, does not produce any entanglement. To create entanglement, we increase the drive power, such that the weak-coupling condition is broken and the counter-rotating-wave (CRW) terms $\propto \delta{c}^\dagger \delta{b}^\dagger + \delta{c} \delta{b}$ in the linearized optomechanical interaction cannot be neglected, which correspond to the PDC interaction yielding the optomechanical entanglement~\cite{Vitali07}.  The entanglement is further distributed to the exciton-phonon and exciton-photon subsystems when the exciton mode resonates with the mechanical Stokes sideband, i.e., $-\Delta_x \approx \omega_b \gg \kappa_x$ (Fig.~\ref{fig2}(a))~\cite{Li18,Zuo}. In this situation, i.e., $\tilde{\Delta}_c \approx -\Delta_x \approx  \omega_b$, all bipartite subsystems are entangled, as shown in Fig~\ref{fig2}(b), and the EOM system shares genuine tripartite entanglement, as witnessed by a nonzero $R^{\rm min}_\tau$ (inset of Fig~\ref{fig2}(b)).  The complementary relation of the dashed curve and the solid and dot-dashed curves in Fig~\ref{fig2}(b) indicates that the exciton-phonon and exciton-photon entanglement are transferred from the photon-phonon entanglement. Similar entanglement distribution phenomena in multipartite systems have been observed in cavity magnomechanics~\cite{Li18} and optomagnomechanics~\cite{Li23}.

The exciton-photon entanglement can be understood straightforwardly as follows. The mechanical motion scatters the driving photons at frequency $\omega_0$ onto the two mechanical sidebands at frequencies $\omega_0 \pm \omega_b$. When the exciton and cavity modes are resonant with the two sidebands, i.e., $\tilde{\Delta}_c \approx -\Delta_x \approx  \omega_b$, the optomechanical Stokes and anti-Stokes scatterings are simultaneously activated and enhanced (here Stokes photons and excitons are of the excitation-exchange interaction).  The Stokes scattering corresponds to the PDC interaction, leading the Stokes photons to be entangled with the mechanical oscillator, whereas the anti-Stokes scattering results in the state-swap (beam-splitter) interaction between the anti-Stokes photons and the oscillator. Therefore, the two sidebands become entangled via the mediation of the mechanical motion.  Since the exciton and cavity modes resonate with the two sidebands, respectively, the exciton and cavity modes thus get entangled.

\begin{figure}[t]
	\includegraphics[width=\linewidth]{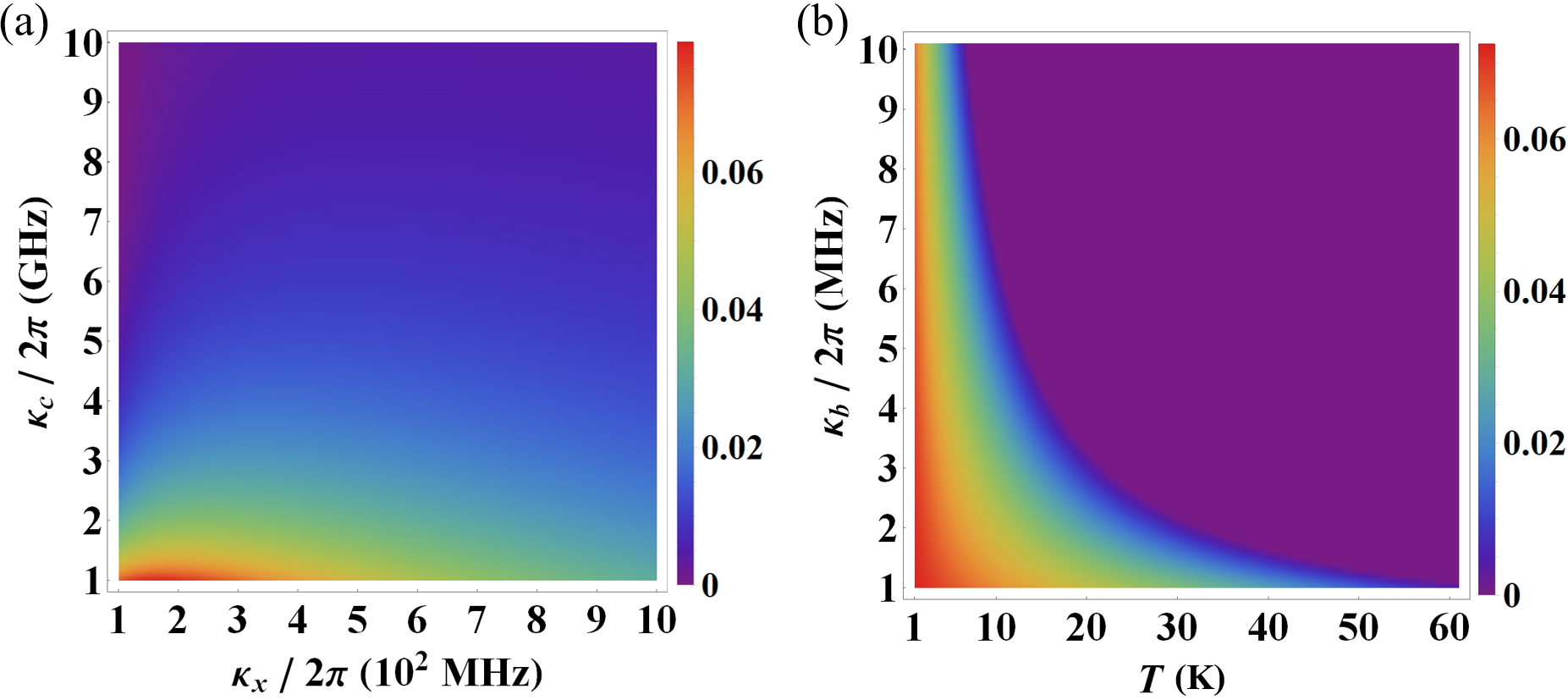}
	\caption{Stationary exciton-photon entanglement versus (a) dissipation rates $\kappa_x$ and $\kappa_c$; (b) bath temperature $T$ and mechanical damping rate $\kappa_b$. We take $\Delta_x = -1.1 \omega_b$, and the other parameters are the same as in Fig~\ref{fig2}(b).}
	\label{fig3}
\end{figure}

In getting Fig~\ref{fig2}(b), we employ the following feasible parameters~\cite{Santos,EOM1,Perrin13}: $\omega_b/2\pi = 20$ GHz, $\kappa_b/2\pi = 1$ MHz, $\kappa_c/2\pi = 1$~GHz, $\kappa_x/2\pi = 10^2$~MHz, $G_0/2\pi = 10$~MHz, $\Omega/2\pi = 6$~THz (corresponding to the drive power $P\,\,{\approx}\,\,26$~mW for $\omega_0/2\pi\,\,{\approx}\,\,345$ THz), and at temperature $T = 1$~K. We take a moderate coupling $g/2\pi = 0.9$~GHz, and the exciton and cavity frequencies are nearly resonant and on both sides of the drive frequency. 
Under these parameters, the mechanical motion is cooled to its quantum ground state with the mean phonon number of $0.03$ at the optimal condition $\tilde{\Delta}_c \approx -\Delta_x \approx  \omega_b$.
In Fig~\ref{fig3}, we study the impact of various dissipation rates of the system and the bath temperature on the exciton-photon entanglement. Clearly, the entanglement is robust against all dissipation rates and is still present for $\kappa_c$, $\kappa_x$, $\kappa_b$ being up to $\sim$10 GHz, 1 GHz, and 10 MHz, respectively, based on the parameters of Fig~\ref{fig2}(b). Due to the relatively high mechanical frequency in the typical EOM system, the entanglement survives at a temperature up to $\sim60$~K.

\section{The Polariton-Mechanics system}\label{model2}

When the exciton-photon system enters the strong-coupling regime, $g>\kappa_c, \kappa_x$, excitons and cavity photons are hybridized forming two exciton polaritons, which are termed as the upper polariton (UP) and the lower polariton (LP), respectively. The theory presented in Secs.~\ref{model1} and~\ref{En1} is based on the original exciton and cavity modes and thus becomes inefficient for the study of the entanglement between two polaritons. In this case, it is more convenient to express the theory in terms of two polaritons. Consequently, we rewrite the EOM Hamiltonian~\eqref{HHH} with the polariton operators
\begin{align}\label{HHHul}
	\begin{split}
		H/\hbar &= \! \omega_u U^\dagger U + \omega_{l} L^\dagger L + \omega_b b^\dagger b		
		\\&+ G_0 \left( b + b^\dagger \right) \Big(U^\dagger U \sin^2\theta+ U^\dagger L \sin\theta \cos\theta 
		\\& \,\,\,\,\,\,\,\,\,\,\,\,\,\,\,\,\,\,\,\,\,\,\,\,\,\,\,\,\,\,\,\,\,\,\, + L^\dagger U \cos\theta \sin\theta + L^\dagger L \cos^2\theta \Big)
		\\&+ i\Omega \Big( U^\dagger \sin\theta e^{-i\omega_0t} + L^\dagger \cos\theta e^{-i\omega_0t}
		\\&\,\,\,\,\,\,\,\,\,\,\,\,\,\, - U \sin\theta e^{i\omega_0t} - L \cos\theta e^{i\omega_0t} \Big) \, ,
	\end{split}
\end{align}
where $U$ and $L$ ($U^\dag$ and $L^\dag$) denote the annihilation (creation) operators of the UP and LP, respectively, which are the hybridization of the exciton and cavity modes via the unitary Hopfield transformation~\cite{Hopfield58}: $U=x \cos\theta + c\sin\theta$ and $L=- x\sin\theta + c\cos\theta$, with the corresponding eigenfrequencies
\begin{align}\label{eigenfreq}
	\begin{split} 
\omega_{u}&=\frac{1}{2}\Big[\omega_x+\omega_c + \sqrt{(\omega_x-\omega_c)^2+4g^2} \Big],  \\
\omega_{l}&=\frac{1}{2}\Big[\omega_x+\omega_c - \sqrt{(\omega_x-\omega_c)^2+4g^2} \Big]. 
	\end{split}
\end{align}
The proportions of excitons and cavity photons in the two polaritons are characterized by the Hopfield coefficients $\sin^2\theta$ and $\cos^2\theta$, where $\theta =\frac{1}{2} \arctan \frac{2g}{\omega_x-\omega_c} \in [0,\frac{\pi}{2}]$, implying that the proportions can be varied by altering the coupling $g$ and/or the exciton-photon detuning $\omega_x-\omega_c$. Therefore, the mechanical mode couples to both the polaritons because both of them contain the photon component, as shown in the Hamiltonian \eqref{HHHul}. When the cavity photons and excitons are resonant, yielding $\theta = \frac{\pi}{4}$, both the UP and LP are exactly half-photon, half-exciton with the minimum frequency splitting $\omega_{u} - \omega_{l} = 2 g$. Since both photons and excitons are bosons, the exciton polaritons, as their linear superposition, are also bosons, and satisfy the bosonic commutation relation $[k,k^\dagger]=1$ ($k = U,L$).

Likewise, we obtain the following QLEs, in the frame rotating at the drive frequency, concerning two polariton modes and a mechanical mode:
\begin{align}
	\begin{split}
		\dot U=&-i\Delta_u U - i G_0 \left( b + b^\dagger \right) \left( U \sin^2\theta + L\sin \theta \cos \theta \right) 
		\\& - \kappa_u U - \delta \kappa L  + \Omega \sin\theta + \sqrt[]{2\kappa_u}U^{in},  \\
		\dot L=&-i\Delta_l L - i G_0 \left( b + b^\dagger \right) \left(L \cos^2\theta + U\cos \theta \sin \theta \right) 
		\\& - \kappa_l L - \delta \kappa U + \Omega \cos\theta + \sqrt[]{2\kappa_l}L^{in}, \\
		\dot{b}=&-i\omega_b b - i G_0 \Big(U^\dagger U \sin^2\theta+U^\dagger L \sin\theta \cos\theta 
		\\&+ L^\dagger U \cos\theta \sin\theta + L^\dagger L \cos^2\theta \Big) - \kappa_b b + \sqrt[]{2\kappa_b} b^{in},
	\end{split}
\end{align}
where $\Delta_{u} = \omega_{u}-\omega_0$ and $\Delta_{l} = \omega_{l}-\omega_0$ denote the polariton-drive detunings; $\kappa_u = \kappa_x \cos^2\theta + \kappa_c \sin^2\theta$ and $\kappa_l = \kappa_x \sin^2\theta + \kappa_c \cos^2\theta $ are the dissipation rates of the two polariton modes; and $\delta \kappa \equiv (\kappa_c-\kappa_x) \sin\theta \cos\theta$ signifies the dissipative coupling between the polaritons stemming from the unbalanced decay rates of the original modes, i.e., $\kappa_x \neq \kappa_c$. $U^{in} \equiv  (\sqrt[]{2\kappa_x} \cos \theta x^{in} + \sqrt[]{2\kappa_c} \sin \theta c^{in}) /\sqrt[]{2\kappa_u}$ and $L^{in} \equiv (- \sqrt[]{2\kappa_x} \sin \theta x^{in} + \sqrt[]{2\kappa_c} \cos \theta c^{in}) /\sqrt[]{2\kappa_l}$ represent the input noises entering the UP and LP, which are the combination of the input noises $x^{in}$ and $c^{in}$ of the exciton and cavity modes.

Similarly, the linearization of the system dynamics around the steady-state values leads to the following QLEs for the quantum fluctuations of the system:
\begin{align}\label{QLEsUL}
	\begin{split}
		\dot{\delta U}{=}&\,{-} \big( i\tilde{\Delta}_u + \kappa_u \big) \delta U {-} \,\big(i G_b  + \delta \kappa \big) \delta L {-} \,G_{u,b} \big( \delta b + \delta b^\dagger \big) {+}\, \sqrt[]{2\kappa_u}U^{in},  \\
		\dot{\delta L}{=}&\,{-} \big( i\tilde{\Delta}_l + \kappa_l \big) \delta L {-} \,\big(i G_b  + \delta \kappa \big) \delta U {-} \,G_{l,b} \big( \delta b + \delta b^\dagger \big) {+}\, \sqrt[]{2\kappa_l}L^{in},  \\
		\dot{\delta b}\,{=}& \,{-} \big( i \omega_b + \kappa_b \big) \delta b   \\ 
		&\, {-} \Big(G_{u,b} \delta U^\dagger + G_{l,b} \delta L^\dagger - G_{u,b}^* \delta U - G_{l,b}^* \delta L \Big) + \sqrt[]{2\kappa_b}b^{in},	
	\end{split}
\end{align}
where $\tilde{\Delta}_u  = \Delta_u + 2 G_0 \langle b \rangle \sin^2\theta$ and $\tilde{\Delta}_l  = \Delta_l + 2 G_0 \langle b \rangle \cos^2\theta$ are the effective polariton-drive detunings, including the frequency shift caused by the optomechanical interaction; $G_{u,b} = \left(G_u \sin\theta + G_l \cos \theta \right) \sin \theta$ and $G_{l,b} = \left(G_u \sin\theta + G_l \cos \theta \right) \cos \theta$ are the coupling strengths between the two polaritons and the mechanical mode, respectively, where $G_{u}= i G_0\langle U \rangle$ and $G_{l}= i G_0\langle L \rangle$; and $G_b = G_0 \langle b \rangle \sin 2\theta$ denotes the coupling between the two polaritons via the mediation of the mechanical mode. The steady-state averages are given by
\begin{align}\label{aveUL}
	\begin{split}
		\langle U \rangle =& \frac{\Omega \left[ \delta \kappa  \cos \theta - i \sin\theta \left( \tilde{\Delta}_l - 2 G_0 \langle b \rangle \cos^2 \theta - i \kappa_l \right) \right]}{\big(\tilde{\Delta}_l - i \kappa_l \big) \big(\tilde{\Delta}_u - i \kappa_u \big) + \delta \kappa^2 - G_b \, \big( G_b - 2 i \delta \kappa \big)  }  ,\\
		\langle L \rangle =& \frac{\Omega \left[ \delta \kappa \sin \theta - i \cos\theta \left( \tilde{\Delta}_u - 2 G_0 \langle b \rangle \sin^2 \theta - i \kappa_u \right) \right]}{\big(\tilde{\Delta}_l - i \kappa_l \big) \big(\tilde{\Delta}_u - i \kappa_u \big) + \delta \kappa^2 - G_b\, \big( G_b - 2 i \delta \kappa \big) }  ,\\
	   \langle b \rangle =& - \frac{G_0}{\omega_b} \big| \langle U \rangle \sin \theta + \langle L \rangle \cos \theta \big|^2.
	\end{split}
\end{align}
Note that since the single-photon optomechanical coupling $G_0$ is typically small~\cite{MA14} and $G_0 \langle b \rangle \propto G_0^2$, one can safely neglect the weak coupling terms $G_b \left(\delta U^\dag \delta L + \delta U \delta L^\dag \right)$ in the QLEs~\eqref{QLEsUL} and the last term of the denominator of $\langle U \rangle$ ($\langle L \rangle$) in Eq.~\eqref{aveUL}, and assume the effective detunings to be $|\tilde{\Delta}_{u,l}| \simeq |\Delta_{u,l}|$, as the optimal detunings for entanglement correspond to $|\Delta_{u,l}| \simeq \omega_b$, which will be shown later.

In the same way, the QLEs~\eqref{QLEsUL} can be rewritten in terms of the quadrature fluctuations $\left(\delta X_u,\delta Y_u,\delta X_l,\delta Y_l,\delta X_b,\delta Y_b \right)$, which can be cast in the matrix form similarly as in Eq.~\eqref{uAn}.  The steady-state CM $V'$, with its entries defined as $V'_{ij}=\frac{1}{2} \langle u'_i(t)u'_j(t^\prime) + u'_j(t^\prime)u'_i(t) \rangle$, where $u'(t)=\left[\delta X_u(t),\delta Y_u(t),\delta X_l(t),\delta Y_l(t),\delta X_b(t),\delta Y_b(t) \right]^{\rm T}$, can be achieved by solving the Lyapunov equation
\begin{align}
	\begin{split}
	{\cal A'} V'+V'{\cal A'}^{\rm T} = -D',
	\end{split}
\end{align}
where the diffusion matrix $D'=\mathrm{Diag} \big[\kappa_u(2N_u {+}\,1),\kappa_u(2N_u {+}\,1),$ $\kappa_l(2N_l +1),\kappa_l(2N_l + 1),\kappa_b(2N_b + 1),\kappa_b(2N_b +1)\big] + \frac{1}{2} \tan2\theta$  $\left[-\kappa_u(2N_u {+}\,1)+\kappa_l(2N_l {+}\,1) \right] \bm{\sigma}^x\,\otimes\,\rm{I}_{2\times2}\,\oplus\,\bm{0}_{2\times2}$, with $\bm{\sigma}^x$ being the $x$-Pauli matrix, 
and $N_u = \frac{1}{2}\big\{ [\kappa_x \cos^2 \theta (2 N_x + 1) + \kappa_c \sin^2 \theta (2 N_c + 1) ]/\kappa_u -1 \big\}$ and $N_l = \frac{1}{2} \big\{ [\kappa_x \sin^2 \theta (2 N_x + 1) + \kappa_c \cos^2 \theta (2 N_c + 1) ]/\kappa_l - 1 \big\}$ as the mean thermal excitation numbers of the UP and LP, respectively. The drift matrix ${\cal A'}$ is given by
\begin{widetext}
\begin{align}
	\cal A'=\begin{pmatrix}
		-\kappa_u & {\Delta}_u & -\delta \kappa & 0 & - 2 {\rm Re}\,G_{u,b} &0 \\
		-{\Delta}_u & -\kappa_u &0 & -\delta \kappa& - 2 {\rm Im}\,G_{u,b} & 0\\
		-\delta \kappa & 0 & -\kappa_l & {\Delta}_l & - 2 {\rm Re}\,G_{l,b} &0 \\
		0 & -\delta \kappa&-{\Delta}_l & -\kappa_l & - 2 {\rm Im}\,G_{l,b} & 0\\
		0 & 0 & 0 & 0 & -\kappa_b & \omega_b \\
		{- 2 {\rm Im}\,G_{u,b}} & 2 {\rm Re}\,G_{u,b} & {- 2 {\rm Im}\,G_{l,b}} & 2 {\rm Re}\,G_{l,b} & -\omega_b & -\kappa_b\\
	\end{pmatrix}.
\end{align}
\end{widetext}
With the CM $V'$ in hand, we can then calculate the bipartite (tripartite) entanglement in the system using the logarithmic negativity (the minimum residual contangle) introduced in Sec.~\ref{model1}.

\section{Entanglement between two Exciton Polaritons}\label{En2}

\begin{figure}[t]
	\includegraphics[width=\linewidth]{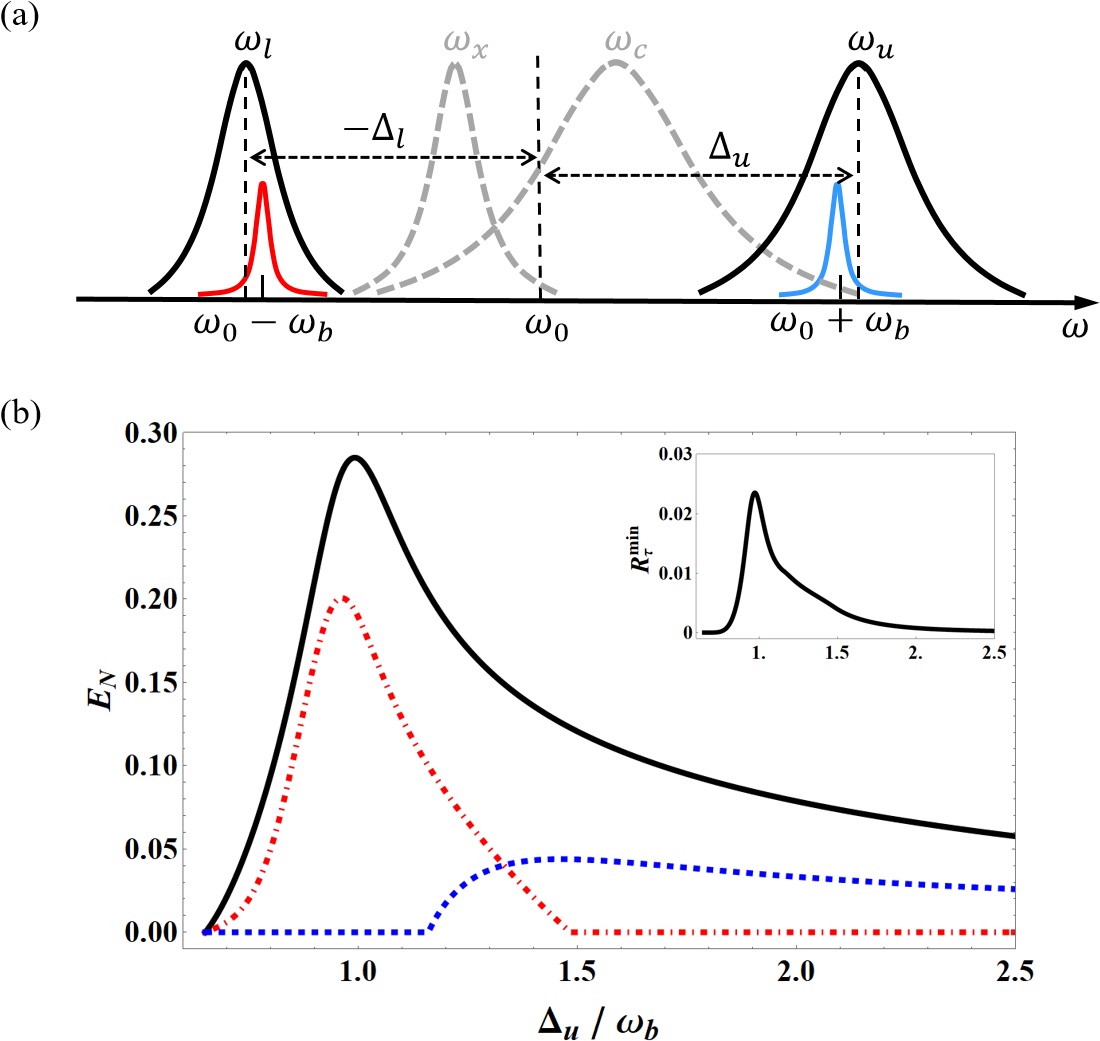}
	\caption{(a) Strongly-coupled excitons and cavity photons form two exciton polaritons at frequencies $\omega_u$ and $\omega_l$. When the UP resonates with the anti-Stokes (blue) sideband and the LP resonates with the Stokes (red) sideband, the system exhibits genuine tripartite entanglement. (b) Stationary UP-LP (solid), LP-phonon (dot-dashed), and UP-phonon (dashed) entanglement and tripartite (inset) entanglement versus detuning $\Delta_u=-\Delta_l$. See text for the other parameters. }
	\label{fig4}
\end{figure}

As analyzed in Sec.~\ref{En1}, the two optomechanical sidebands at $\omega_0 \pm \omega_b$ are entangled due to the mediation of the mechanical oscillator involved in both the Stokes and anti-Stokes scatterings. It is quite natural to conjecture that the entanglement between the two polaritons can be achieved when they are respectively resonant with the two sidebands, i.e., $\Delta_u = - \Delta_l = \omega_b$ (Fig~\ref{fig4}(a)).  Since both the polaritons contain the photon component and their exciton component has no coupling with the mechanical mode, the interaction between {\it each} polariton and the mechanics is fully determined by the interaction between its photon component and the mechanics, i.e., the dispersive optomechanical interaction~\cite{Zuo}. The UP (LP) being resonant with the anti-Stokes (Stokes) sideband greatly enhances the strength of the anti-Stokes (Stokes) scattering and thus the optomechanical cooling (PDC) interaction. A unique feature and advantage of the polariton system is that the strength of the cooling (PDC) interaction associated with the UP (LP) is adjustable by changing the weight of the photon component in the polaritons via altering $\theta$.  As also discussed in Sec.~\ref{En1}, the emergence of entanglement in the system requires the combination of both the cooling and PDC interactions for eliminating thermal noise and creating entanglement, respectively.  The fact that the cavity photons enter the two polaritons with the weights of $\sin^2 \theta$ and $\cos^2 \theta$, respectively, implies that there is a trade-off between the cooling and PDC interactions for achieving the maximal entanglement, which indicates an optimal value of $\theta$.

To reach such an optimal $\theta$ in a real experiment, we fix the exciton frequency and the exciton-photon coupling, e.g., at $\omega_x/2\pi = 345$ THz and $g/2\pi = 13$ GHz ($g<\omega_b$), but set the cavity frequency $\omega_c$ as a variable, which can be realized by adopting the configuration where the microcavity layer thickness is tapered by growth and thus the cavity resonance can be continuously tuned across the sample~\cite{Deng10,Weisbuch92}. This leads to a continuous varying of $\theta$, since $\theta =\frac{1}{2} \arctan \frac{2g}{\omega_x-\omega_c}$. It should be noted that as the cavity frequency changes, the two polariton-drive detunings $\Delta_u$ and $|\Delta_l|$ change, but they are generally not equal (cf., Eq.~\eqref{eigenfreq}).  To maintain the optimal condition $\Delta_u = -\Delta_l \approx \omega_b$ as the cavity frequency varies, we set the drive frequency to be $\omega_c$-dependent, i.e., $\omega_0 = \frac{\omega_x + \omega_c}{2}$.

In Fig.~\ref{fig4}(b), we present the results of the bipartite and tripartite entanglement of the system.  As expected, the entanglement between the two polaritons is maximized, $E_N\approx0.28$, at the detunings $\Delta_u = - \Delta_l \approx \omega_b$, and the LP-phonon entanglement is also optimized at this point, because the LP-phonon entanglement is a direct result of the Stokes scattering, of which the strength is maximal when the LP resonates with the Stokes sideband.  However, at this point the UP-phonon entanglement is absent. This is because the effective interaction between the UP and the mechanical mode is solely the cooling (beam-splitter) interaction, which does not yield entanglement. One can increase the UP-mechanics coupling strength by raising the weight of the photon component in the UP, such that, as analyzed in Sec.~\ref{En1}, the weak-coupling condition is broken and the CRW terms start to play the role in creating the UP-phonon entanglement.   
The increase of $\Delta_u$ in Fig.~\ref{fig4}(b) corresponds to a continuous rise of $\omega_c$, which leads to a continuous increase of $\theta$, starting from $\theta=\frac{\pi}{4}$ corresponding to $\omega_c=\omega_x$. This results in a growing weight of the photon component in the UP and thus an increasing UP-mechanics coupling strength. When $\Delta_u>\approx1.2 \omega_b$, the coupling strength becomes sufficiently strong, leading to the emergence of the UP-phonon entanglement. A much larger detuning $\Delta_u$ causes the polaritons to deviate from the mechanical sidebands, thus diminishing the entanglement. The system also exhibits genuine tripartite entanglement around $\Delta_u= -\Delta_l \approx \omega_b$, as shown in the inset of Fig~\ref{fig4}(b).  In getting Fig.~\ref{fig4}(b), we keep the coupling strength $G_l$ fixed at $|G_l |/2\pi = 0.6$ GHz by adjusting the drive power as the detuning $\Delta_u$ varies (cf. Eq.~\eqref{aveUL}). The corresponding $\Omega/2\pi=3.5$ THz and drive power $P\approx8.6$ mW at the optimal detunings $\Delta_u = - \Delta_l = \omega_b$. The other parameters are the same as those in Fig~\ref{fig2}.

\begin{figure}[t]
	\includegraphics[width=\linewidth]{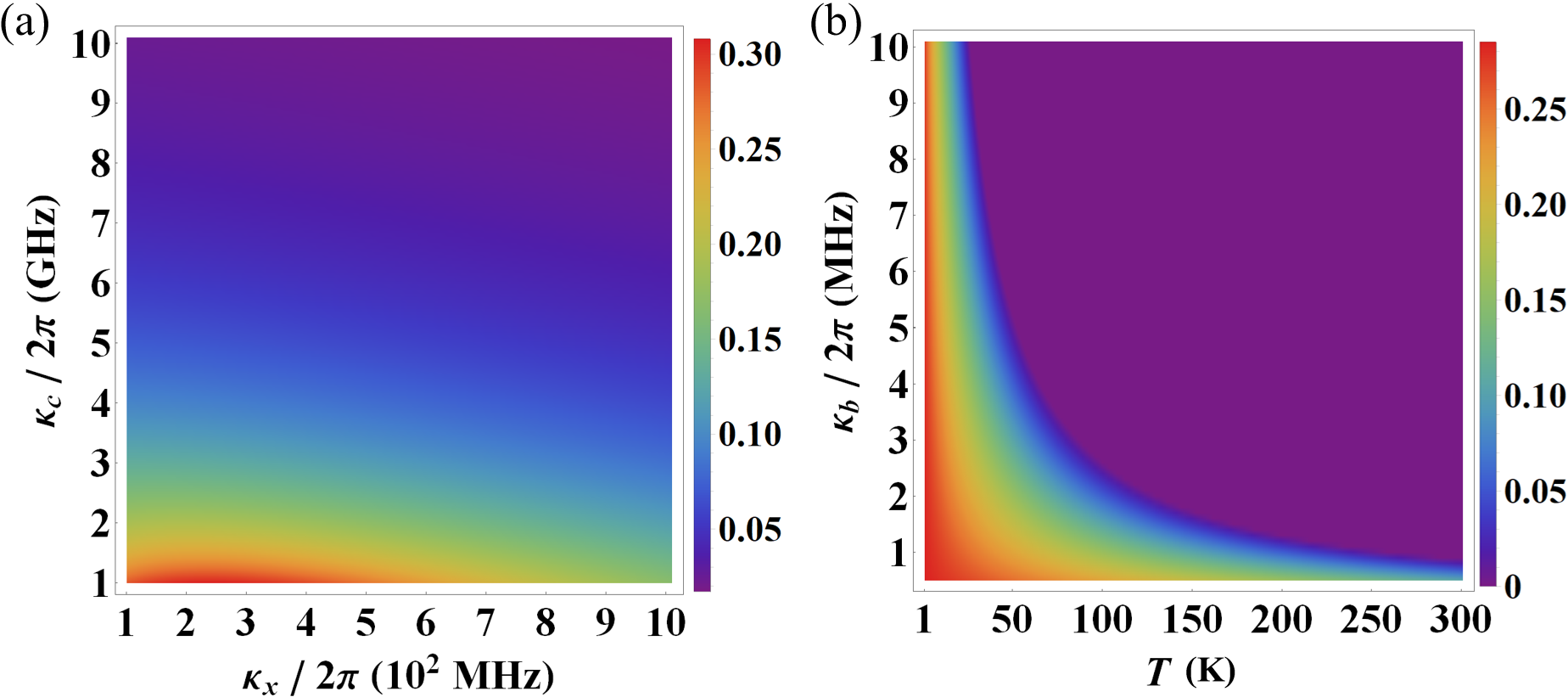}
	\caption{Stationary entanglement between two polaritons versus (a) $\kappa_x$ and $\kappa_c$; (b) bath temperature $T$ and $\kappa_b$. We take the optimal detunings $\Delta_u = - \Delta_l = \omega_b$. The other parameters are the same as in Fig~\ref{fig4}(b).}
	\label{fig5}
\end{figure}

The entanglement of two polaritons is much stronger than the exciton-photon entanglement generated in Sec.~\ref{En1} even with a smaller drive power. The entanglement is stationary and robust with respect to the dissipation rates of the three modes, as seen in Fig.~\ref{fig5}. The entanglement can be achieved even at room temperature if using a sub-MHz mechanical damping rate. For example, the entanglement is still present for the temperature up to $300$ K with $\kappa_b/2\pi=0.5$ MHz (Fig.~\ref{fig5}(b)), corresponding to a mechanical $Q$ factor of $4\times10^4$, which is high but reachable~\cite{Perrin13,Santos}.

\section{Discussion and Conclusion}\label{conc}

In view of the presented two entanglement protocols in the weak- and strong-coupling regimes, they share similarities but also have their own characteristics.  In both situations, the generation of entanglement requires the combination of the cooling and PDC interactions. For the cooling interaction, in both cases it is achieved by activating the optomechanical anti-Stokes scattering, realized by keeping the anti-Stokes sideband resonant with the cavity mode in the weak-coupling regime (with the UP in the strong-coupling regime). However, their entanglement mechanisms are not exactly the same. In entangling excitons and cavity photons, the power of the red-detuned drive field must be sufficiently strong to activate the CRW terms to have the PDC interaction (Sec.~\ref{En1}); whereas in the strong-coupling case of entangling two polaritons, the PDC interaction can be directly provided by the optomechanical Stokes scattering associated with the LP (Sec.~\ref{En2}), {\it which does not require a strong drive field}. The fact that the cavity photons enter the two polaritons to simultaneously participate in both the optomechanical Stokes and anti-Stokes scatterings makes the protocol in the strong-coupling case more efficient, reflected by the fact that the entanglement between the two polaritons is much stronger than the exciton-photon entanglement even under a weaker drive field.

The generated bipartite and tripartite entanglement can be verified by measuring the corresponding CMs~\cite{Vitali07,Li18}. For the exciton-photon entanglement, the cavity field quadratures can be measured directly by homodyning the cavity output field. The excitonic quadratures can be detected by sending a weak probe field that realizes a state-swap interaction with the excitons, and by homodyning the cavity output field of the probe field. Note that in typical exciton-microcavity systems, the cavity decay rate is much larger, such that when the laser drive is switched off and all cavity photons dissipate, the excitonic state remains practically unchanged, at which time the probe field is sent. Similarly, to measure the entanglement of two polaritons, one can successively send two weak probe fields that are respectively resonant with the two polaritons, and perform the homodyne detection of the cavity output fields. To verify the tripartite entanglement, the mechanical state must be accessed. This can be realized by sending a weak red-detuned light into the cavity, which activates an effective optomechanical state-swap interaction and maps the mechanical state to the cavity field, and by homodying the cavity output field. Due to the much longer mechanical coherence time, the probe light can be sent after all cavity photons and excitons die out.

In conclusion, we present a systemic theory for entangling excitons and microcavity photons, or two exciton polaritons when they are strongly coupled. The idea is to introduce a dispersively coupled mechanical mode into the exciton-photon system, which brings in the optomechanical cooling and PDC interactions, responsible for eliminating thermal noise and creating entanglement in the system, respectively. By appropriately adjusting the strengths of the two interactions, stationary exciton-photon or polariton-polariton entanglement can be achieved. Impressively, room-temperature polariton entanglement can potentially be obtained by improving relevant experimental parameters of the EOM system. 

Although many classical phenomena have been studied in the EOM system~\cite{EOM1,EOM2,Fainstein20,Santos21,Santos23}, quantum effects have been rarely explored~\cite{Santos,EOM3}. This work represents the first entanglement study, to our knowledge, in the field of EOM.
The work also provides theoretical guidance for the experimental realization of the exciton-photon entanglement and the polariton entanglement, which may find applications in quantum information processing with exciton polaritons, e.g., entangled polaritons can lead to the emission of frequency-entangled photon pairs~\cite{Ciuti04}.

\section*{ACKNOWLEDGMENTS}

This work has been supported by National Key Research and Development Program of China (Grant no. 2022YFA1405200) and National Natural Science Foundation of China (Grant no. 92265202).


\begin{thebibliography}{99}



\bibitem{Deng10}
H. Deng, H. Haug, and Y. Yamamoto, Exciton-polariton Bose-Einstein condensation, Rev. Mod. Phys. {\bf 82}, 1489 (2010).

\bibitem{Keeling07}
J Keeling, F M Marchetti, M H Szymańska, and P B Littlewood, Collective coherence in planar semiconductor microcavities, Semicond. Sci. Technol. {\bf 22}, R1 (2007).

\bibitem{Hopfield58}
J. J. Hopfield, Theory of the contribution of excitons to the complex dielectric constant of crystals, Phys. Rev. {\bf 112}, 1555 (1958).

\bibitem{Weisbuch92}
C. Weisbuch, M. Nishioka, A. Ishikawa, and Y. Arakawa, Observation of the coupled exciton-photon mode splitting in a semiconductor quantum microcavity, Phys. Rev. Lett. {\bf 69}, 3314 (1992).

\bibitem{Houdre94}
R. Houdré, C. Weisbuch, R. P. Stanley, U. Oesterle, P. Pellandini, and M. Ilegems, Measurement of cavity-polariton dispersion curve from angle-resolved photoluminescence experiments, Phys. Rev. Lett. {\bf 73}, 2043 (1994).



\bibitem{Khitrova06}
G. Khitrova, H. M. Gibbs, M. Kira, S. W. Koch, and A. Scherer, Vacuum Rabi splitting in semiconductors, Nat. Phys. {\bf 2}, 81 (2006).

\bibitem{Sun08}
L. Sun, Z. Chen, Q. Ren, K. Yu, L. Bai, W. Zhou, H. Xiong, Z. Q. Zhu, and X. Shen, Direct observation of whispering gallery mode polaritons and their dispersion in a ZnO tapered microcavity, Phys. Rev. Lett. {\bf 100}, 156403 (2008).

\bibitem{Lidzey98}
D. G. Lidzey, D. D. C. Bradley, M. S. Skolnick, T. Virgili, S. Walker, and D. M. Whittaker, Strong exciton-photon coupling in an organic semiconductor microcavity, Nature {\bf 395}, 53 (1998).

\bibitem{Kena-Cohen08}
S Kéna-Cohen, M. Davanço, and S. R. Forrest, Strong exciton-photon coupling in an organic single crystal microcavity, Phys. Rev. Lett. {\bf 101}, 116401 (2008).

\bibitem{Su21}
R. Su, A. Fieramosca, Q. Zhang, H. S. Nguyen, E. Deleporte, Z. Chen, D. Sanvitto, T. C. H. Liew, and Q. Xiong, Perovskite semiconductors for room-temperature exciton-polaritonics, Nat. Mater. {\bf 20}, 1315 (2021).





\bibitem{Cerna13}
R. Cerna, Y. Léger, T. K. Paraïso, M. Wouters, F. Morier-Genoud, M. T. Portella-Oberli, and B. Deveaud, Ultrafast tristable spin memory of a coherent polariton gas, Nat. Commun. {\bf 4}, 2008 (2013).

\bibitem{Tsintzos08}
S. I. Tsintzos, N. T. Pelekanos, G. Konstantinidis, Z. Hatzopoulos, and P. G. Savvidis, A GaAs polariton light-emitting diode operating near room temperature, Nature {\bf 453}, 372 (2008).

\bibitem{Chakraborty19}
J. Gu, B. Chakraborty, M. Khatoniar, and V. M. Menon, A room-temperature polariton light-emitting diode based on monolayer WS$_2$, Nat. Nanotechnol. {\bf 14}, 1024 (2019).

\bibitem{Ballarini13}
D. Ballarini, M. D. Giorgi, E. Cancellieri, R. Houdré, E. Giacobino, R. Cingolani, A. Bramati, G. Gigli, and D. Sanvitto, All-optical polariton transistor, Nat. Commun. {\bf 4}, 1778 (2013).

\bibitem{Zasedatelev19}
A. V. Zasedatelev, A. V. Baranikov, D. Urbonas, F. Scafirimuto, U. Scherf, T. Stöferle, R. F. Mahrt, and P. G. Lagoudakis, A room-temperature organic polariton transistor, Nat. Photonics {\bf 13}, 378 (2019).



\bibitem{BEC1}
J. Kasprzak, M. Richard, S. Kundermann, A. Baas, P. Jeambrun, J. M. J. Keeling, F. M. Marchetti, M. H. Szymańska, R. André, J. L. Staehli, V. Savona, P. B. Littlewood, B. Deveaud, and Le Si Dang, Bose–Einstein condensation of exciton polaritons, Nature {\bf 443}, 409 (2006).

\bibitem{BEC2}
R. Balili, V. Hartwell, D. Snoke, L. Pfeiffer, and K. West, Bose-Einstein condensation of microcavity polaritons in a trap, Science {\bf 316}, 1007 (2007).

\bibitem{Bogoliubov}
S. Utsunomiya, L. Tian, G. Roumpos, C. W. Lai, N. Kumada, T. Fujisawa, M. Kuwata-Gonokami, A. Löffler, S. Höfling, A. Forchel, and Y. Yamamoto, Observation of Bogoliubov excitations in exciton-polariton condensates, Nat. Phys. {\bf 4}, 700 (2008).

\bibitem{vortices}
K. G. Lagoudakis, M. Wouters, M. Richard, A. Baas, I. Carusotto, R. André, Le Si Dang, and B. Deveaud-Plédran, Quantized vortices in an exciton–polariton condensate, Nat. Phys. {\bf 4}, 706 (2008).  

\bibitem{fluid}
A. Amo, D. Sanvitto, F. P. Laussy, D. Ballarini, E. del Valle, M. D. Martin, A. Lemaître, J. Bloch, D. N. Krizhanovskii, M. S. Skolnick, C. Tejedor, and L. Viña, Collective fluid dynamics of a polariton condensate in a semiconductor microcavity, Nature {\bf 457}, 291 (2009). 



\bibitem{Imamoglu96}
A. Imamoglu, R. J. Ram, S. Pau, and Y. Yamamoto, Nonequilibrium condensates and lasers without inversion: Exciton-polariton lasers, Phys. Rev. A {\bf 53}, 4250 (1996).

\bibitem{Malpuech02}
G. Malpuech, A. D. Carlo, A. Kavokin, J. J. Baumberg, M. Zamfirescu, and P. Lugli, Room-temperature polariton lasers based on GaN microcavities, Appl. Phys. Lett. {\bf 81}, 412 (2002).

\bibitem{Deng03}
H. Deng, G. Weihs, D. Snoke, J. Bloch, and Y. Yamamoto, Polariton lasing vs. photon lasing in a semiconductor microcavity, Proc. Natl. Acad. Sci. {\bf 100}, 15318 (2003).

\bibitem{Christopoulos07}
S. Christopoulos, G. Baldassarri Höger von Högersthal, A. J. D. Grundy, P. G. Lagoudakis, A. V. Kavokin, J. J. Baumberg, G. Christmann, R. Butté, E. Feltin, J.-F. Carlin, and N. Grandjean, Room-temperature polariton lasing in semiconductor microcavities, Phys. Rev. Lett. {\bf 98}, 126405 (2007).

\bibitem{Bajoni08}
D. Bajoni, P. Senellart, E. Wertz, I. Sagnes, A. Miard, A. Lemaître, and J. Bloch, Polariton laser using single micropillar GaAs-GaAlAs semiconductor cavities, Phys. Rev. Lett. {\bf 100}, 047401 (2008).

\bibitem{Kena-Cohen10}
S. Kéna-Cohen and S. R. Forrest, Room-temperature polariton lasing in an organic single-crystal microcavity, Nat. Photonics {\bf 4}, 371 (2010).

\bibitem{Schneider13}
C. Schneider, A. Rahimi-Iman, N. Y. Kim, J. Fischer, I. G. Savenko, M. Amthor, M. Lermer, A. Wolf, L. Worschech, V. D. Kulakovskii, I. A. Shelykh, M. Kamp, S. Reitzenstein, A. Forchel, Y. Yamamoto, and S. Höfling, An electrically pumped polariton laser, Nature {\bf 497}, 348 (2013).

\bibitem{Bhattacharya13}
P. Bhattacharya, B. Xiao, A. Das, S. Bhowmick, and J. Heo, Solid state electrically injected exciton-polariton laser, Phys. Rev. Lett. {\bf 110}, 206403 (2013).

\bibitem{Bhattacharya14}
P. Bhattacharya, T. Frost, S. Deshpande, M. Z. Baten, A. Hazari, and A. Das, Room temperature electrically injected polariton laser, Phys. Rev. Lett. {\bf 112}, 236802 (2014).



\bibitem{Ciuti04}
C. Ciuti, Branch-entangled polariton pairs in planar microcavities and photonic wires, Phys. Rev. B {\bf 69}, 245304 (2004).

\bibitem{Liew18}
T. C. H. Liew and Y. G. Rubo, Quantum exciton-polariton networks through inverse four-wave mixing, Phys. Rev. B {\bf 97}, 041302(R) (2018).

\bibitem{DS}
D. Stefanatos and E. Paspalakis, Efficient entanglement generation between exciton-polaritons using shortcuts to adiabaticity, Opt. Lett. {\bf 43}, 3313 (2018). 

\bibitem{Feng}
J. Feng, H. Li, Z. Sun, and T. Byrnes, Entanglement generation and detection in split exciton-polariton condensates, Phys. Rev. A {\bf 108}, 053301 (2023).

\bibitem{Cuevas}
\'A. Cuevas {\it et al.}, First observation of the quantized exciton-polariton field and effect of interactions on a single polariton, Sci. Adv. {\bf 4}, eaao6814 (2018).


\bibitem{Boulier14}
T. Boulier, M. Bamba, A. Amo, C. Adrados, A. Lemaitre, E. Galopin, I. Sagnes, J. Bloch, C. Ciuti, E. Giacobino, and A. Bramati, Polariton-generated intensity squeezing in semiconductor micropillars, Nat. Commun. {\bf 5}, 3260 (2014).

 

\bibitem{Santos}
P. V. Santos, and A. Fainstein, Polaromechanics: polaritonics meets optomechanics, Opt. Mater. Express  {\bf 13}, 1974 (2023).

\bibitem{EOM1}
O. Kyriienko, T. C. H. Liew, and I. A. Shelykh, Optomechanics with cavity polaritons: Dissipative coupling and unconventional bistability, Phys. Rev. Lett. {\bf 112}, 076402 (2014).

\bibitem{EOM2}
B. Jusserand, A. N. Poddubny, A. V. Poshakinskiy, A. Fainstein, and A. Lemaitre, Polariton resonances for ultrastrong coupling cavity optomechanics in 
GaAs/AlAs multiple quantum wells, Phys. Rev. Lett. {\bf 115}, 267402 (2015).

\bibitem{Fainstein20}
D. L. Chafatinos, A. S. Kuznetsov, S. Anguiano, A. E. Bruchhausen, A. A. Reynoso, K. Biermann, P. V. Santos, and A. Fainstein, Polariton-driven phonon laser, Nat. Commun. {\bf 11}, 4552 (2020).

\bibitem{Santos21}
A. S. Kuznetsov, D. H. O. Machado, K. Biermann, and P. V. Santos, Electrically Driven Microcavity Exciton-Polariton Optomechanics at 20 GHz, Phys. Rev. X {\bf 11}, 021020 (2021).

\bibitem{EOM3}
N. Carlon Zambon, Z. Denis, R. De Oliveira, S. Ravets, C. Ciuti, I. Favero, and J. Bloch, Enhanced cavity optomechanics with quantum-well exciton polaritons, Phys. Rev. Lett. {\bf 129}, 093603 (2022).

\bibitem{Santos23}
A. S. Kuznetsov, K. Biermann, A. Reynoso, A. Fainstein, and  P. V. Santos, Microcavity phonoritons---a coherent optical-to-microwave interface, Nat. Commun. {\bf 14}, 5470 (2023). 




\bibitem{Perrin13}
A. Fainstein, N. D. Lanzillotti-Kimura, B. Jusserand, and B. Perrin, Strong optical-mechanical coupling in a vertical GaAs/AlAs microcavity for subterahertz phonons and near-infrared light, Phys. Rev. Lett. {\bf 110}, 037403 (2013).

\bibitem{MA14}
M. Aspelmeyer, T. J. Kippenberg, and F. Marquardt, Cavity optomechanics, Rev. Mod. Phys. {\bf 86}, 1391 (2014).


\bibitem{Zoller}
C. W. Gardiner and P. Zoller, {\it Quantum Noise} (Springer, Berlin, 2000).

\bibitem{Parks}
P. C. Parks and V. Hahn, {\it Stability Theory} (Prentice Hall, New York, 1993).

\bibitem{Vitali07}
D. Vitali, S. Gigan, A. Ferreira, H. R. Bohm, P. Tombesi, A. Guerreiro, V. Vedral, A. Zeilinger, and M. Aspelmeyer, Optomechanical entanglement between a movable mirror and a cavity field, Phys. Rev. Lett. {\bf 98}, 030405 (2007).


\bibitem{Ent}
G. Vidal and R. F. Werner, Computable measure of entanglement, Phys. Rev. A {\bf 65}, 032314 (2002); G. Adesso, A. Serafini, and F. Illuminati, Extremal entanglement and mixedness in continuous variable systems, Phys. Rev. A {\bf 70}, 022318 (2004); M. B. Plenio, Logarithmic negativity: A full entanglement monotone that is not convex, Phys. Rev. Lett. {\bf 95}, 090503 (2005).

{\bibitem{Simon}
R. Simon, Peres-Horodecki separability criterion for continuous variable systems, Phys. Rev. Lett. {\bf 84}, 2726 (2000).}

\bibitem{Cont}
G. Adesso and F. Illuminati, Entanglement in continuous-variable systems: recent advances and current perspectives, J. Phys. A {\bf 40}, 7821 (2007). 

{\bibitem{Adesso}
G. Adesso and F. Illuminati, Continuous variable tangle, monogamy inequality, and entanglement sharing in Gaussian states of continuous variable systems, New J. Phys. {\bf 8}, 15 (2006).}

{\bibitem{Coffman}
V. Coffman, J. Kundu, and W. K. Wootters, Distributed entanglement, Phys. Rev. A {\bf 61}, 052306 (2000).
}

\bibitem{DV08}
C. Genes, A. Mari, P. Tombesi, and D. Vitali, Robust entanglement of a micromechanical resonator with output optical fields, Phys. Rev. A {\bf 78}, 032316 (2008).

\bibitem{Li18}
J. Li, S.-Y. Zhu, and G. S. Agarwal, Magnon-photon-phonon entanglement in cavity magnomechanics, Phys. Rev. Lett. {\bf 121}, 203601 (2018).


\bibitem{Zuo}
X. Zuo, Z.-Y. Fan, H. Qian, M.-S. Ding, H. Tan, H. Xiong, and J. Li, Cavity magnomechanics: from classical to quantum, New J. Phys. {\bf 26}, 031201 (2024).

\bibitem{Li23}
Z.-Y. Fan, L. Qiu, S. Gr\"oblacher, and J. Li, Microwave-optics Entanglement via Cavity Optomagnomechanics, Laser Photonics Rev. 2200866 (2023); arXiv:2208.10703.




\end{thebibliography}
\end{document}